\documentclass[a4paper,10p]{article}
\usepackage{epsfig}
\usepackage{amssymb}
\begin{document}

\title{EVOLUTION OF THE TRUNCATED MELLIN MOMENTS OF THE PARTON DISTRIBUTIONS
IN QCD ANALYSIS}

\author{Dorota Kotlorz\footnote{Opole University of Technology, Division of
Physics, Ozimska 75, 45-370 Opole, Poland, e-mail: 
{\tt d.strozik-kotlorz@po.opole.pl}} and Andrzej Kotlorz\footnote{Opole
University of Technology, Division of Mathematics and Applied Informatics,
Luboszycka 3, 45-036 Opole, Poland, e-mail: {\tt a.kotlorz@po.opole.pl}}}
\date{October 23, 2013}
\maketitle

\abstract{
We review evolution equations for the truncated Mellin moments of the
parton distributions and some their applications in QCD analysis.
The main finding of the presented approach is that the
$n$th truncated moment of the parton distribution obeys
also the DGLAP equation but with a rescaled splitting 
function $P'(z)=z^n P(z)$. This allows one to avoid the
problem of dealing with the experimentally unexplored  
Bjorken-$x$ region. The evolution equations for truncated
moments are universal - they are valid in each order of
perturbation expansion and can be useful additional tool
in analysis of unpolarized as well as polarized
nucleon structure functions.\\\\
PACS{12.38.Bx 11.55.Hx}
}

\section{Introduction}
\label{Intro}

Understanding the details of the Bjorken-$x$ and $Q^2$ dependence of the
nucleon structure functions is one of the most important challenges in
high energy physics. Perturbative QCD provides a comprehensive framework for
describing deep inelastic lepton-hadron scattering (DIS) and hadron-hadron
collisions (H-H) in current and planned experiments. In this framework, where
the high energy collisions proceed via partonic constituents of the hadron,
a key role is played by universal parton distribution functions (PDFs).
According to the factorization theorem, DIS or H-H cross section is
a convolution of a short distance interaction, described by the
partonic cross section $\hat{\sigma}_f$ and a long distance structure
described by the parton distribution~$q_f$.

Particularly interesting is a study of polarized processes which
provides knowledge about the spin structure of the nucleon. Though recent
experimental data and NLO analyses suggest that valence quarks carry the
expected fraction of the nucleon spin, the main questions are still open:
how the nucleon spin is distributed among its constituents: quarks
(particularly sea quarks with negative helicity) and gluons and how the
dynamics of these constituent interactions depends on spin. New experimental
data in the resonance region from Jefferson Lab together with complementary
data from HERMES, COMPASS and RHIC, are a crucial step towards better
understanding of not only the flavor decomposition and gluon contributions
to the nucleon spin but also the quark-hadron duality. The main goal of
present polarized experiments is to determine the nucleon spin structure
functions $g_1(x,Q^2)$, $g_2(x,Q^2)$ and their moments which are essential
in testing QCD sum rules.
The theoretical approach usually used in the description of these experimental
results are the QCD evolution equations for the parton densities
which change with $Q^2$ according to the well-known DGLAP equations
\cite{b1,b2,b3,b4}.
This standard DGLAP approach operates on the parton densities $q$; hence their
moments, which are, e.g., the contributions to the proton spin and other
sum rules, can be obtained by the integration of the parton densities $q$
over Bjorken-$x$.

Alternatively, one can directly study the $Q^2$ evolution of the Mellin
moments of the parton densities.
The moments provide a natural framework in QCD analysis, as they
originate from OPE - the basic formalism of quantum field theory.
The idea of truncated Mellin moments (TMM) of the parton densities in QCD
analysis was introduced and developed in the late 1990s
\cite{b5,b6,b7,b8}.
The authors obtained the non-diagonal differential evolution
equations in which the $n$th truncated moment coupled to all higher ones.
The evolution equations for TMM within the $ln^2x$ approximation were found in
\cite{b9}, and finally DGLAP-type diagonal integro-differential evolution
equations for the single and double truncated moments of the parton
distribution functions were derived in \cite{b10}, \cite{b11} and
\cite{b12}.
Evolution equations for double truncated moments and their application to
study the quark-hadron duality were also discussed in \cite{b13}.

The main finding of the truncated moments approach is that the $n$th moment
of the parton distribution obeys also the DGLAP equation but with a rescaled
splitting function $P'(z)=z^n P(z)$ \cite{b10}.
This approach allows one to restrict the analysis to the experimentally
available Bjorken-$x$ region.
The evolution equations for TMM are universal - they are valid
in each order of perturbation expansion (then Wilson coefficients
rescale in the same way as the splitting functions) and can be an additional
tool in QCD analysis of the unpolarized as well as the polarized nucleon
structure functions.

In this paper, we review our main results on the truncated moments
approach. In the next section, we present the evolution equations for the
truncated moments of the parton distributions. Section 3 contains 
relations between the truncated and untruncated Mellin moments, useful for
solving evolution equations. Section 4 describes determination of the parton
densities from their truncated moments.
Implications of the TMM approach for analysis of the polarized structure
functions $g_1$ and $g_2$ are presented in Sections 5 and 6, respectively.
We show the evolution of TMM of $g_1$ together with the predictions for the
Bjorken sum rule. Then we present the Wandzura-Wilczek relation and sum
rules in terms of the truncated moments. We also give the evolution
equations. Finally, in summary, we highlight possible future applications
of the truncated moments approach in QCD analysis. 

\section{The evolution equations for the truncated Mellin moments
of the parton distributions}
\label{sec.2}

The structure functions of the nucleon can be expressed in terms of the parton
distributions. These depend on two kinematic variables: the Bjorken $x$ and
$Q^2=-q^2$ with $q$ being the four-momentum transfer in the deep-inelastic
lepton-nucleon scattering (DIS). The scaling variable is defined as
$x=Q^2/(2pq)$, where $p$ is the nucleon four-momentum. The strong interactions
between quarks and gluons cause changes in the parton densities.
For medium and large $x$, the $Q^2$ evolution of the parton distributions is
described by the standard Dokshitzer-Gribov-Lipatov-Altarelli-Parisi (DGLAP)
equations \cite{b1}-\cite{b4}:
\begin{equation}\label{eq.2.1}
\frac{dq(x,Q^2)}{d\ln Q^2}=\frac{\alpha_s(Q^2)}{2\pi}\; (P\otimes q)(x,Q^2),
\end{equation}
where $\alpha_s(Q^2)$ is the running coupling, $\otimes$ denotes the Mellin
convolution
\begin{equation}\label{eq.2.3}
(A\otimes B)(x)\equiv\int\limits_{x}^{1} \frac{dz}{z}\,
A\left(\frac{x}{z}\right)\,B(z),
\end{equation}
and $P(z)$ is the splitting function, which can be expanded in a power series
of $\alpha_s(Q^2)$.

In the DGLAP approach, the main role is played by the PDFs and in our TMM
approach we study directly the $Q^2$ evolution of the truncated moments
of the PDFs.
In \cite{b10}, we found that the single truncated moments of the
parton distributions $q(x,Q^2)$, defined as
\begin{equation}\label{eq.2.8}
\bar{q}^{n}(x_0,Q^2)=\int\limits_{x_0}^1 dx\, x^{n-1}\, q(x,Q^2),
\end{equation}
obey the DGLAP-like equation
\begin{equation}\label{eq.2.9}
\frac{d\bar{q}^n(x_0,Q^2)}{d\ln Q^2}=
\frac{\alpha_s(Q^2)}{2\pi}\; (P'\otimes \bar{q}^n)(x_0,Q^2).
\end{equation}
A role of the splitting function is played here by $P'(n,z)$:
\begin{equation}\label{eq.2.10}
P'(n,z)= z^n\, P(z).
\end{equation}
Since the experimental data cover only a limited range of $x$, except very
small $x\rightarrow 0$ as well as large $x\rightarrow 1$, it is very natural
and convenient to deal with the double truncated moments.
Truncation at large $x$ is less important in comparison to the small-$x$
limit because of the rapid decrease of the parton densities as
$x\rightarrow 1$; nevertheless, a comprehensive theoretical analysis requires
an equal treatment of both truncated limits.

The double truncated moment
\begin{equation}\label{eq.2.11}
\bar{q}^{n}(x_{min},x_{max},Q^2)=
\int\limits_{x_{min}}^{x_{max}} dx\, x^{n-1}\, q(x,Q^2),
\end{equation}
as it is a subtraction of two single truncated ones,
also satisfies the DGLAP-type evolution Eq.~(\ref{eq.2.9}) \cite{b11,b12,b13}:
\begin{equation}\label{eq.2.12}
\frac{d\bar{q}^n(x_{min},x_{max},Q^2)}{d\ln Q^2} =
\frac{\alpha_s(Q^2)}{2\pi}\;
\int\limits_{x_{min}}^{1}\frac{dz}{z}\; P'(n,z)\;
\bar{q}^n\left( \frac{x_{min}}{z}, \frac{x_{max}}{z}, Q^2 \right)
\end{equation}
with $P'$ given again by Eq.~(\ref{eq.2.10}).

Our approach, Eq.~(\ref{eq.2.9}) - Eq.~(\ref{eq.2.12}), is valid for the
coupled DGLAP equations for quarks and gluons and for any approximation
(LO, NLO, NNLO, etc.). For clarity, we present here only the nonsinglet
and leading order part. In higher order analysis (eg. NLO), truncated
moments of the structure functions assume the form
\begin{eqnarray}\label{eq.2.13}
\bar g_{1\,n}(x,Q^2) = \frac{1}{2}\sum_q e_q^2\;\times \nonumber\\
\times\left [\Delta\bar{q}_{\,n}(x,Q^2) + \frac{\alpha_s(Q^2)}{2\pi}
\left ( C_q'(n)\otimes\Delta\bar{q}_{\,n} + C_G'(n)\otimes
\Delta\bar{G}_{\,n}\right )(x,Q^2) \right ],
\end{eqnarray}
where the Wilson coefficients rescale in the same way as the splitting
functions:
\begin{equation}\label{eq.2.14}
C_{i}'(n,x) = x^n\,C_{i}(x).
\end{equation}
Let us emphasize that the evolution equations for the double
truncated moments, Eq.~(\ref{eq.2.12}),
are in fact a valuable generalization of those for the single truncated and
untruncated ones. Setting $x_{min} = x_0$ or $x_{min} = 0$ and $x_{max} = 1$
one obtains Eq.~(\ref{eq.2.9}) or the well-known renorm-group equation for
the moments
\begin{equation}\label{eq.2.15}
\frac{d\bar{q}^{n}(Q^2)}{d\ln Q^2}=\frac{\alpha_s(Q^2)}{2\pi}\;
\gamma^n(Q^2)\, \bar{q}^{n}(Q^2),
\end{equation}
respectively.

In the next section, we present relations between truncated and
untruncated moments which are useful for solving the evolution equations.

\section{Relations between truncated and untruncated moments}
\label{sec.3}

The evolution equations for the truncated moments, Eq.~(\ref{eq.2.9}), are
very similar to those for the PDF. In both cases, one deals with
functions of two variables $x$ and $Q^2$ (with additionally fixed index $n$
for moments), which obey the differentio-integral Volterra-like equations.
The only difference lies in the splitting function, which for moments has the
rescaled form Eq.~(\ref{eq.2.10}). This similarity allows one to solve the
equations for truncated moments with the use of standard methods of solving the
DGLAP equations. Analysis of the evolution performed in moment space,
when applied to the truncated moments, implies dealing with such an exotic
structure as `Moment of Moment'. Let us discuss this in detail and introduce
some useful relations involving untruncated and truncated Mellin moments.

There are in literature several methods for the solution of the
integro-differential DGLAP equations. They are based on either the polynomial
expansion or the Mellin transformation - for review see, e.g., \cite{b14}.
In our previous studies on the evolution of the truncated moments we used the
Chebyshev polynomial technique \cite{b15}, earlier widely applied by Jan
Kwieci\'nski in many QCD treatments - for details see, e.g., Appendix of
\cite{b16}. Using this method one obtains the system of linear differential
equations instead of the original integro-differential ones. The Chebyshev
expansion provides a robust method of discretising a continuous problem.

An alternative approach is based on the Mellin transformation and the
moment factorization. Taking the $s$-th moment of the evolution equation
(\ref{eq.2.9}) one obtains
\begin{equation}\label{eq.4.1}
\frac{dM^{s,\, n}(Q^2)}{d\ln Q^2}=\frac{\alpha_s(Q^2)}{2\pi}\;
\gamma^{s+n}(Q^2)\, M^{s,\, n}(Q^2),
\end{equation}
where $M^{s,\, n}$ denotes here the $s$-th (untruncated) moment of the $n-$th
truncated moment of the parton density:
\begin{equation}\label{eq.4.2}
M^{s,\, n}(Q^2) = \int\limits_{0}^{1} dx\, x^{s-1}\, \bar{q}^{n}(x,Q^2).
\end{equation}
Analogically to the well known solutions for the PDFs, we can immediately
write down solutions for the truncated moments:
\begin{equation}\label{eq.4.3}
M^{s,\, n}(Q^2) = M^{s,\, n}(Q_0^2)
\left[\frac{\alpha_s(Q_0^2)}{\alpha_s(Q^2)}\right]^{\,b\,\gamma^{s+n}}
\end{equation}
and
\begin{equation}\label{eq.4.4}
\bar{q}^n(x,Q^2)=\frac{1}{2\pi i}\int\limits_{c-i\infty}^{c+i\infty}
ds\, x^{-s}\, M^{s,\, n}(Q^2).
\end{equation}
The quantity $M^{s,\, n}$, which is rather exotic and has no physical meaning,
can be replaced by the usual truncated moment $\bar{q}$.
In \cite{b12}, we found useful relations between the truncated and untruncated
moments, namely:
\begin{equation}\label{eq.4.5}
M^{s,\, n} =
\frac{1}{s}\int\limits_{0}^1 dz\, z^{s+n-1} q(z) = 
\frac{1}{s}\, \bar{q}^{s+n},
\end{equation}
\begin{equation}\label{eq.4.6}
\bar{q}^n(x,Q^2)=\frac{1}{2\pi i}\int\limits_{c-i\infty}^{c+i\infty}
ds\,\frac{x^{-s}}{s}\, \bar{q}^{s+n}(Q^2)
\end{equation}
and
\begin{equation}\label{eq.4.7}
\bar{q}^s(Q^2) = (s-n)\, M^{s-n,\, n}(Q^2) = 
(s-n)\int\limits_{0}^{1} dx\, x^{s-n-1}\,\bar{q}^n(x,Q^2).
\end{equation}
Eqs. (\ref{eq.4.5}), (\ref{eq.4.6}) and (\ref{eq.4.7}) have a large
practical meaning in solving the evolution equations of TMM.
Particularly, Eq.~(\ref{eq.4.6}) seems to be helpful when the untruncated
moments are known, e.g., from lattice calculations.

In the next sections we will discuss some implications of our approach for
QCD analysis.

\section{Determination of the parton distributions from their TMM}
\label{sec.4}

The TMM approach, which refers to the physical values - moments (not to the
parton densities), allows one to study directly their evolution and
the scaling violation. The solutions for truncated moments can also be used
in the determination of the parton distribution functions via differentiation
\begin{equation}\label{eq.6.1}
q(x,Q^2) = -x^{1-n}\:\frac{\partial\bar{q}_{n}(x,Q^2)}{\partial x},
\end{equation}
In order to reconstruct initial parton densities at scale $Q_0^2$ from their
truncated moments, given, e.g., by experimental data at scale $Q^2$,
we evolve moments between these two scales down (from $Q^2$ to $Q_0^2$) and
then perform the final fit of free parameters - for details see \cite{b11}.

We proceed with the following steps:\\
1. Preparing available experimental data for moments $\bar{q}_{n}(x_0,Q_1^2)$
as a function of $x_{min}\leq x_0\leq 1$ at the same scale $Q_1^2$.\\
2. Interpolation of the given data points into the points that are Chebyshev
nodes. This allows us to use the Chebyshev polynomial technique for solving
the evolution equations.\\
3. Evolution of the truncated moments from $Q_1^2$ to $Q_2^2$, according to
(\ref{eq.2.1}), for different $x_{min}\leq x_0\leq 1$.\\
4. Reconstruction of the parton density $q(x,Q_2^2)$ from its truncated moment
at the same scale $Q_2^2$ by applying the minimizing algorithm to fit free
parameters.\\
In \cite{b11}, we tested this method on the nonsinglet function parametrized
in the general form
\begin{equation}\label{eq.6.2}
q(x,Q_0^2) = N(\alpha, \beta, \gamma)\:x^{\alpha} (1-x)^{\beta}(1+\gamma x),
\end{equation}
and also on the original fits for HERMES \cite{b17} and COMPASS \cite{b18}
data.
%***********
\begin{figure}[ht]
\begin{center}
\includegraphics[width=85mm]{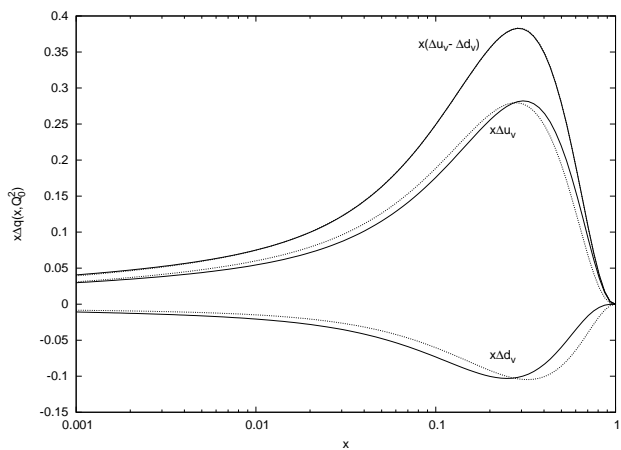}
\caption{Initial spin-dependent valence quark distributions
$x(\Delta u_{v}-\Delta d_{v})$, $x\Delta u_{v}$ and $x\Delta d_{v}$ at
$Q_0^2=4\;{\rm GeV}^2$:
dotted - reconstructed from HERMES data for the first truncated moment of
the nonsinglet polarized function $g_1^{NS}$ at $Q^2=5\;{\rm GeV}^2$
\cite{b17}, solid - original BB fit \cite{b19}.
Plots for $x(\Delta u_{v}-\Delta d_{v})$ overlap each other.}
\end{center}
\end{figure}
\begin{figure}[ht]
\begin{center}
\includegraphics[width=85mm]{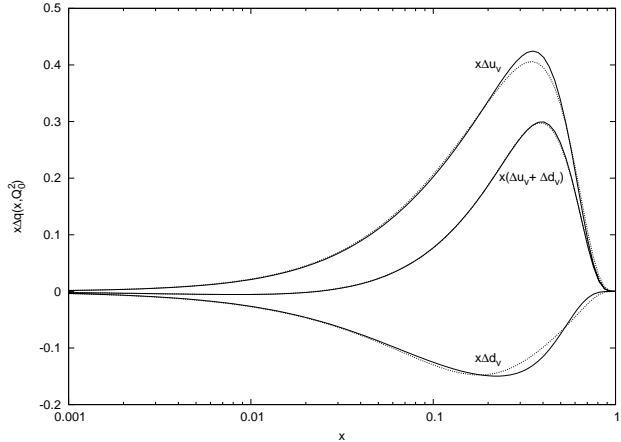}
\caption{Initial spin-dependent valence quark distributions
$x(\Delta u_{v}+\Delta d_{v})$, $x\Delta u_{v}$ and $x\Delta d_{v}$ at
$Q_0^2=0.5\;{\rm GeV}^2$:
dotted - reconstructed from COMPASS data for the first truncated moment of
the function $\Delta u_{v}+\Delta d_{v}$ at $Q^2=10\;{\rm GeV}^2$
\cite{b18}, solid - original DNS fit \cite{b20}.
Plots for $x(\Delta u_{v}+\Delta d_{v})$ overlap each other.}
\end{center}
\end{figure}
In Figs. 1-2, we show the spin-dependent valence quark distributions
reconstructed from HERMES \cite{b17} and COMPASS \cite{b18} data.
From these plots one can see satisfactory agreement between the reconstructed
fits and experimental data. The reconstructed combined functions
$x(\Delta u_{v}-\Delta d_{v})$ and $x(\Delta u_{v}+\Delta d_{v})$ overlap
HERMES and COMPASS results, respectively. For the extracted valence quark
densities alone the agreement is worse but still acceptable. We have found,
however, that these fits are not unique and equally good agreement with the
data can be obtained with the use of other (not only BB and DNS respectively)
sets of free parameters. When the number of adjustable parameters is large
($>3$) and there are no experimental points from the low-$x$
region $x<0.001$, one cannot distinguish which fit is the best one. Only an
additional constraint for small-$x$ behaviour of the parton densities makes
the fit procedure more reliable. Let us also mention that due to its large-$x$
sensitivity, the second moment can be helpful in the precise final
reconstruction of the parton density \cite{b11}.
Concluding, even for the large number of adjustable parameters (6 for HERMES
and 8 for COMPASS data), the presented method of reconstruction can be
a hopeful tool for determining parton densities from experimental results
for their truncated moments. 

\section{TMM in analysis of the spin structure function $g_1$}
\label{sec.5}

For a complete description of the nucleon spin, one needs two polarized
structure functions: $g_1$ and $g_2$. Recently, a new generation of
experiments with high polarized luminosity performed at the Jefferson Lab
allows a more precise study of the polarized structure functions and their
moments. This is crucial in our understanding of the QCD spin sum rules,
higher-twist effects and quark-hadron duality.

The function $g_1$ has a simple interpretation in the parton model:
\begin{equation}\label{eq.5.1}
g_1(x) = \frac{1}{2}\sum_i e_i\, \Delta q_i(x),
\end{equation}
Powerful tools to study the internal spin structure of the nucleon are
sum rules. One of them is the Bjorken sum rule (BSR) \cite{BSR} which
refers to the first moment of the nonsinglet spin dependent structure
function $g_1^{NS}(x,Q^2)$:
\begin{equation}\label{eq.5.2}
\Gamma_1^{p-n} = \int_0^1 dx\, g_1^{NS}(x,Q^2)=\int_0^1 dx\, (g_1^p-g_1^n).
\end{equation}
Due to of $SU_f(2)$ flavour symmetry, BSR is regarded as exact. Thus, all of
estimations of polarized parton distributions should be performed under the
assumption that the BSR is valid. In the  limit of the infinite momentum
transfer $Q^2$, BSR has the form
\begin{equation}\label{eq.5.3}
\Gamma_1^{p-n} =  
\int\limits_{0}^{1} dx\, g_1^{NS}(x,Q ^2) =
\frac {1}{6}\left | {\frac{g_A}{g_V}}\right |
\end{equation}
where $|{\frac{g_A}{g_V}}|$ is the neutron $\beta$-decay constant
\begin{equation}\label{eq.5.4}
\left | {\frac{g_A}{g_V}} \right | = F + D = 1.28.
\end{equation}
With the next perturbative orders and higher twists corrections,
the BSR reads
\begin{equation}\label{eq.5.5}
\Gamma_1^{p-n}(Q^2) =
\underbrace{\frac {1}{6} \left | {\frac{g_A}{g_V}} \right | 
\left [\, 1-\frac{\alpha_s}{\pi}-3.58\,\frac{\alpha_s^2}{\pi^2}
-20.21\,\frac{\alpha_s^3}{\pi^3}+\;...\,\right ]}_{leading\;\;\; twist} + 
\underbrace{\sum_{i=2}^{\infty}\frac{\mu_{2i}(Q^2)}{Q^{2i-2}}}_{higher\;\;\;
twists}.
\end{equation}
Since experimental data cover only a restricted region of the Bjorken-$x$
variable and, in fact, provide knowledge on truncated moments of the structure
functions, it is very natural to use the TMM approach in analysis of these data.
The TMM approach allows for a direct study of the contributions to the BSR.
In Figs. 3-5 and in Table 1, we present the results for the evolution of the
first truncated moment of $g_1^{NS}$, namely
\begin{equation}\label{eq.5.6}
\Gamma_1^{p-n}(x_0,1,Q^2) = \int\limits_{x_0}^{1}g_1^{NS}(x,Q^2)\,dx.
\end{equation}
%***
\begin{figure}[ht]
\begin{center}
\includegraphics[width=90mm]{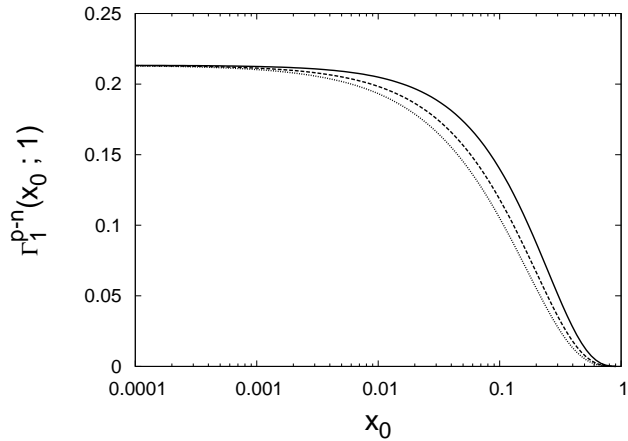}
\caption{First truncated moment of $g_1^{p-n}$ vs truncation point $x_0$
for different $Q^2$: $\rm{1\, GeV^2}$ (solid), $\rm{10\, GeV^2}$ (dashed)
and $\rm{100\, GeV^2}$ (dotted). We assume Regge input:
$g_1^{p-n}(x,Q_0^2) = N(1-x)^3$.} 
\end{center}
\end{figure}
\begin{figure}[ht]
\begin{center}
\includegraphics[width=90mm]{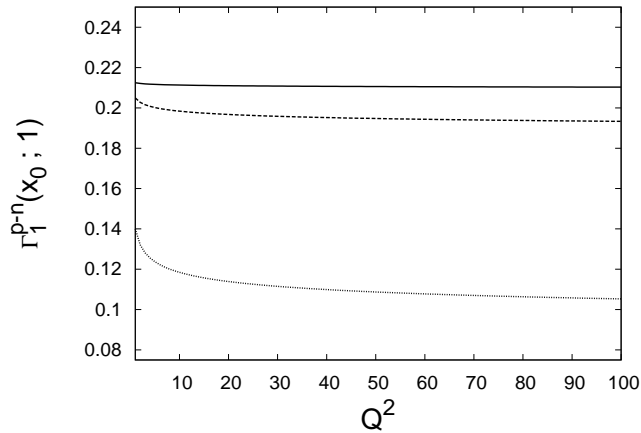}
\caption{First truncated moment of $g_1^{p-n}$ vs evolution scale $Q^2$
for different truncation points $x_0:$ $\rm{0.001}$ (solid),
$\rm{0.01}$ (dashed) and $\rm{0.1}$ (dotted). Regge input parametrization
$g_1^{p-n}(x,Q_0^2) = N(1-x)^3$.}
\end{center}
\end{figure}
\begin{figure}[ht]
\begin{center}
\includegraphics[width=90mm]{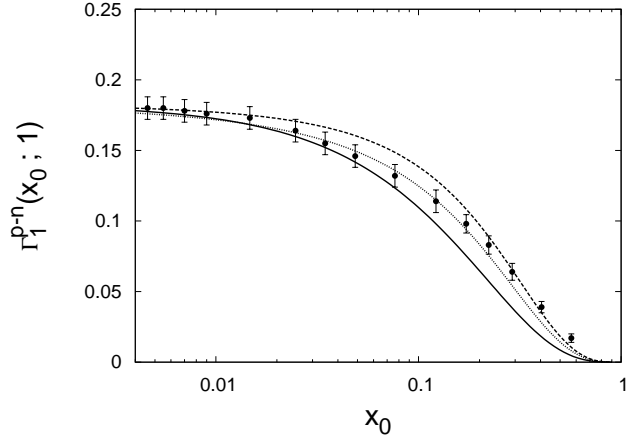}
\caption{First truncated moment of $g_1^{p-n}$ vs truncation point $x_0$ at
evolution scale $Q^2=\rm{3\, GeV^2}$ for different small-$x$ behaviour of
the input parametrization $g_1^{p-n}(x,Q_0^2) = N\,x^{\alpha}(1-x)^3$:
$\alpha =$ $\rm{0}$ (solid), $\rm{-0.2}$ (dashed) and $\rm{-0.4}$ (dotted).
Comparison with COMPASS data \cite{COMPASS}.}
\end{center}
\end{figure}
Figure 3 shows $\Gamma_1^{p-n}(x_0,1,Q^2)$ as a function of the truncation point
$x_0$ for different scales $Q^2$. Figure 4 presents the $Q^2$ evolution of
$\Gamma_1^{p-n}(x_0,1,Q^2)$ for different $x_0$. In Fig.~5 we compare our
predictions for different input parametrization of the nonsinglet structure
function $g_1^{NS}(x,Q_0^2)$ at the initial scale $Q_0^2=1\;{\rm GeV}^2$:
\begin{equation}\label{eq.5.7}
g_1^{NS}(x,Q_0^2) = N x^{\alpha}(1-x)^{\beta}
\end{equation}
with experimental COMPASS data \cite{COMPASS}. Here we take into account
the first perturbative correction to the BSR: $-\alpha_s /\pi$.
In Table 1 we present truncated contributions to the Bjorken sum rule obtained
for different small-$x$ behaviour of the input parametrizations of
$g_1^{NS}(x,Q_0^2)$. Our predictions are compared with experimental HERMES
\cite{b17} and COMPASS \cite{COMPASS} data.
%\begin{footnotesize}
\begin{table}[ht]
%\title{Table 1}
\begin{center}
\begin{tabular}{|c|c|c|c|c|}
\hline
INPUT          & $x$ -range & $Q^2\,[\rm{GeV}^2]$ & $\Gamma_1^{NS}$ & EXP $\Gamma_1^{NS}$\\ \hline
$(1-x)^3$ &                    &   & 0.161 & HERMES \\
$x^{-0.2}(1-x)^3$ & 0.021-0.9  & 5 & 0.149 & $0.1479\pm 0.0055\pm 0.0142$\\
$x^{-0.4}(1-x)^3$ &            &   & 0.131 & \\ \hline
$(1-x)^3$ &                    &   & 0.177 & COMPASS \\
$x^{-0.2}(1-x)^3$ & 0.004-0.7  & 3 & 0.173 & $0.175\pm 0.009\pm 0.015$\\
$x^{-0.4}(1-x)^3$ &            &   & 0.163 & \\ \hline
\end{tabular}
\caption{Theoretical predictions for truncated contributions to the BSR
for different small-$x$ behaviour of $g_1^{NS}$. Comparison with the
HERMES and COMPASS data.}
\end{center}
\end{table}
%\end{footnotesize}

\section{TMM in analysis of the spin structure function $g_2$}
\label{sec.6}

Unlikely $g_1$, the structure function $g_2$ has no simple interpretation in
the parton model. Due to the technical difficulties of obtaining transversely
polarized targets, the structure function $g_2$ has not been a topic of
investigations for a long time. Recently, new experimental data at low and
intermediate momentum transfers have made $g_2$ also a valuable and hopeful
tool to study the spin structure of the nucleon.
The function $g_2$ provides knowledge on higher twist effects which are
a reflection of the quark-gluon correlations in the nucleon.

The experimental value of the function $g_2$, measured in the small to
intermediate $Q^2$ region, consists of two parts: the twist-2 (leading) and
the higher twist term:
\begin{equation}\label{eq.6.4}
g_2(x,Q^2) = g_2^{LT}(x,Q^2) + g_2^{HT}(x,Q^2).
\end{equation}
The leading-twist term $g_2^{LT}$ can be determined from the structure
function $g_1$ via the Wandzura-Wilczek relation \cite{b23}
\begin{equation}\label{eq.6.5}
g_2^{LT}(x,Q^2) = g_2^{WW}(x,Q^2) = -g_1(x,Q^2) + \int_x^1 \frac{dy}{y}\,
g_1(y,Q^2).
\end{equation}
Then, from the measurements of $g_1$ and $g_2$, using the Wandzura-Wilczek
approximation, Eq.~(\ref{eq.6.5}), one is able to extract the higher-twist
term $g_2^{HT}$.
In \cite{b12}, we found a generalization of the Wandzura-Wilczek relation
in terms of the truncated moments:
\begin{equation}\label{eq.6.7}
\bar{g}_2^n(x_0,Q^2) = \frac{1-n}{n}\: \bar{g}_1^n(x_0,Q^2)
- \frac{x_0^n}{n}\:\bar{g}_1^0(x_0,Q^2),
\end{equation}
where
\begin{equation}\label{eq.6.8}
\bar{g}_{1,2}^{n}(Q^2)=\int\limits_{0}^1 dx\, x^{n-1}\, g_{1,2}(x,Q^2),
\end{equation}
\begin{equation}\label{eq.6.9}
\bar{g}_{1,2}^{n}(x_0,Q^2)=\int\limits_{x_0}^1 dx\, x^{n-1}\, g_{1,2}(x,Q^2),
\end{equation}
and
\begin{equation}\label{eq.6.10}
\bar{g}_1^{0}(x_0,Q^2)=\int\limits_{x_0}^1 \frac{dx}{x}\, g_1(x,Q^2).
\end{equation}
It is easy to see that for the untruncated moments, Eq.~(\ref{eq.6.7}), takes
the well-known form
\begin{equation}\label{eq.6.6}
\bar{g}_2^n(Q^2) = \frac{1-n}{n}\: \bar{g}_1^n(Q^2). 
\end{equation}
From  Eq.~(\ref{eq.6.7}), setting $n=1$ and $x_0\rightarrow 0$, one can
automatically obtain the Burkhardt-Cottingham sum rule (BC) \cite{b24} for
$g_2^{WW}$:
\begin{equation}\label{eq.6.15}
\int\limits_0^1 dx\, g_2(x,Q^2) = 0.
\end{equation}
Using the generalization of the Wandzura-Wilczek relation,
Eq.~(\ref{eq.6.7}), for $n=1$ at two different points of the truncation
and applying the BC sum rule, Eq.~(\ref{eq.6.15}), we obtain an interesting
relation:
\begin{equation}\label{eq.6.16}
\int\limits_{x_1}^{x_2} dx\, g_2^{WW}(x,Q^2) =
(x_2-x_1)\int\limits_{x_2}^1\frac{dx}{x}\,g_1(x,Q^2)-
x_1\int\limits_{x_1}^{x_2}\frac{dx}{x}\, g_1(x,Q^2),
\end{equation}
which can be very useful in determination of the partial twist-2
contribution to the BC sum rule. Namely, setting $x_1=0$ and $x_2=x_0$,
when $x_0\rightarrow 0$, one can get the small-$x$ contribution to the BC
sum rule:  
\begin{equation}\label{eq.6.17}
\int\limits_{0}^{x_0} dx\, g_2^{WW}(x,Q^2) =
x_0\int\limits_{x_0}^1\frac{dx}{x}\,g_1(x,Q^2).
\end{equation}
Now we would like to discuss the problem of the $Q^2$ evolution of $g_2$
\cite{b12}.
While a general DGLAP-type equation for $g_2$ does not exist, for the twist-3
component of $g_2$ suitable evolution equation is known (see eg. \cite{b25}).
In the leading twist-2 approximation, the $Q^2$ evolution of $g_2$ is
governed by the evolution of $g_1$, according to the Wandzura-Wilczek relation.
Since the second term on the r.h.s. of Eq.~(\ref{eq.6.5}) is the $n=0$th
truncated moment of the function $g_1$ Eq.~(\ref{eq.6.10}), we can rewrite
the Wandzura-Wilczek relation in the form
\begin{equation}\label{eq.6.21}
g_2^{WW}(x,Q^2) = -g_1(z,Q^2) + \bar{g}_1^0(z,Q^2)
\end{equation}
and obtain the evolution equation for $g_2^{WW}$:
\begin{equation}\label{eq.6.19}
\frac{dg_2^{WW}(x,Q^2)}{d\ln Q^2} = -\frac{dg_1(x,Q^2)}{d\ln Q^2}+
\frac{d\bar{g}_1^0(x,Q^2)}{d\ln Q^2}.
\end{equation}
It is worth noting that according to Eqs.~(\ref{eq.2.9}),(\ref{eq.2.10}),
the $n=0$th truncated moment of the parton distribution $q$ evolves in the
same way as $q$ itself ($P'(0,z)=P(z)$). Taking this into account in the
case of $g_1$ we obtain from Eqs.~(\ref{eq.6.19}), (\ref{eq.6.21}) the
evolution equation
\begin{equation}\label{eq.6.20}
\frac{dg_2^{WW}(x,Q^2)}{d\ln Q^2} = \frac{\alpha_s(Q^2)}{2\pi}
\int\limits_x^1 \frac{dz}{z}\,P\left(\frac{x}{z}\right)\left[
\bar{g}_1^0(z,Q^2)-g_1(z,Q^2)\right]
\end{equation}
or finally
\begin{equation}\label{eq.6.22}
\frac{dg_2^{WW}(x,Q^2)}{d\ln Q^2} = \frac{\alpha_s(Q^2)}{2\pi}
\int\limits_x^1 \frac{dz}{z}\,P\left(\frac{x}{z}\right)\, g_2^{WW}(z,Q^2).
\end{equation}
The above formula shows that the twist-2 component of the function $g_2$
obeys the standard DGLAP evolution with the same evolution kernel
as $g_1$. In this way, we obtained a system of evolution equations
for
\begin{equation}\label{eq.6.23}
g_2 = g_2^{EXP} = g_2^{WW} + g_2^{twist-3}:
\end{equation}
\begin{equation}\label{eq.6.24}
\frac{d\left[g_2^{EXP}(x,Q^2)-g_2^{WW}(x,Q^2)\right]}{d\ln Q^2} =
\frac{\alpha_s(Q^2)}{2\pi}\int\limits_x^1 \frac{dz}{z}\,P^{twist-3}
\left(\frac{x}{z}\right)\, \left[g_2^{EXP}(z,Q^2)-g_2^{WW}(z,Q^2)\right],
\end{equation}
\begin{equation}\label{eq.6.25}
\frac{dg_2^{WW}(x,Q^2)}{d\ln Q^2} = \frac{\alpha_s(Q^2)}{2\pi}
\int\limits_x^1 \frac{dz}{z}\,P\left(\frac{x}{z}\right)\, g_2^{WW}(z,Q^2).
\end{equation}
%*** 
\begin{figure}[ht]
\begin{center}
\includegraphics[width=90mm]{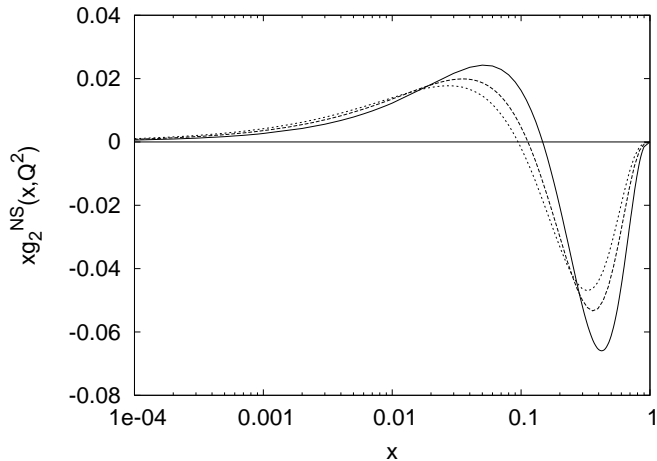}
\caption{
The nonsinglet LO contributions to the polarized structure function
$xg_2^{NS}(x,Q^2)$ as a function of $x$ for different $Q^2$: $\rm{1\, GeV^2}$
(solid), $\rm{10\, GeV^2}$ (dashed) and $\rm{100\, GeV^2}$ (dotted).
Low-$x$ behaviour of $g_1^{NS}\sim x^{-0.4}$.
}
\end{center}
\end{figure}
\begin{figure}[ht]
\begin{center}
\includegraphics[width=90mm]{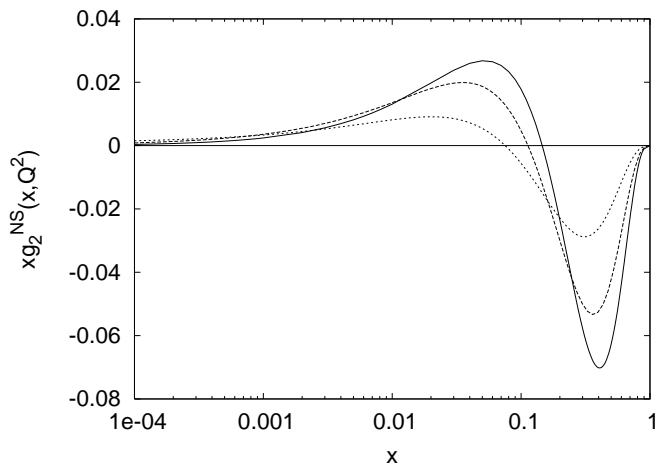}
\caption{
The nonsinglet LO contributions to the polarized structure function
$xg_2^{NS}(x,Q^2)$ at $Q^2=\rm{10\,GeV^2}$ vs $x$ for different
low-$x$ behaviour of $g_1^{NS}\sim x^{\alpha}$: $\alpha = 0$ (solid),
$\alpha = -0.4 $ (dashed) and $\alpha = -0.8$ (dotted).
}
\end{center}
\end{figure}
In Figs.~6-7, we present numerical solutions of Eq.~(\ref{eq.6.22})
calculated in LO, for different low-$x$ behaviour of $g_1^{NS}\sim x^{\alpha}$
and for different $Q^2$.
Fig.~6 shows the predictions for $xg_2^{NS}$ vs $x$ for different scales of
$Q^2$: 1, 10 and 100 $\rm{GeV^2}$.
In the input parametrization of $g_1^{NS}\sim x^{\alpha}$ we assume at the
scale $Q_0^2$ $\alpha = -0.4$. Note that $xg_2^{NS}$ is positive for low-$x$,
at about $x=0.1-0.2$ changes sign and becomes negative for larger $x$.
This is in agreement with the BC sum rule. One can see also that with
increasing $Q^2$, an $x$-intercept of $g_2^{NS}$ occurs at smaller values
of $x$.
In Fig.~7, we compare the predictions for $g_2^{NS}$ for different small-$x$
behaviour of the $g_1^{NS}$ parametrization: $\alpha =0,\,-0.4,\,-0.8$.
We find that more singular small-$x$ behaviour of $g_1$ implies smaller
value of the $x$-intercept of $g_2$.
%***

\section{Summary}
\label{sec.7}

This paper is a review of our studies on the truncated Mellin moments
of the parton distributions.
We presented the evolution equations for the single and double
truncated $n$-th moments and useful relations between the truncated and
untruncated moments. We gave examples of application of our approach to
the determination of the PDFs and to QCD analysis of the spin-dependent
structure functions $g_1$ and $g_2$. We presented the Wandzura-Wilczek
relation in terms of the truncated moments, which implies the truncated
sum rules. We also discussed the system of the DGLAP evolution equations
for $g_2$. We presented the numerical predictions for the evolution of
the truncated moments of $g_1^{NS}$ and the contributions to the BSR.
We tested different small-$x$ behaviour of the initial
parametrization of $g_1^{NS}$ and compared our results with COMPASS data.

The method of the truncated moments enables one direct, efficient study
of the evolution of the moments (and hence sum rules) for non-spin as well
as for spin-dependent parton distributions and can be used in all orders
of perturbative theory. The adaptation of the evolution equations for the
available experimentally $x$-region provides a new, additional tool for
analysis of the nucleon structure functions.

Finally, let us list possible future applications of the TMM approach:
\begin{itemize}
\item
Study of the fundamental properties of the nucleon structure, concerning
moments of $F_1$, $F_2$ and $g_1$. These are: the momentum fraction carried
by quarks, quark helicities contributions to the spin of nucleon and,
what is particularly important, estimation of the polarized gluon
contribution $\Delta G$ from COMPASS and RHIC data.
\item
Determination of Higher Twist (HT) effects from the moments of $g_2$ in the
restricted $x$-region, which will be measured at JLab and can provide
information on the quark-hadron duality.
\item
Test of Burkhardt-Cottingham and Efremov-Leader-Teryaev sum rules \cite{b26},
also for their truncated contributions together with comparison to
experimental data.
\item
Predictions for the generalized parton distributions (GPDs). Moments of the
GPDs can be related to the total angular momentum (spin and orbital) carried
by various quark flavors. Measurements of DVCS, sensitive to GPDs, will be
carried out at JLab. This would be an important step towards a full
understanding of the nucleon spin.
\end{itemize}

Concluding, in light of the recent progress in experimental program,
theoretical efforts in improving our knowledge of the nucleon
structure functions and their moments are of great importance.\\
\\
%*****

\section*{Acknowledgements}

DK is grateful to JINR for an opportunity to work here and to BLTP for
the warm hospitality. 

% Bibliography
\begin{small}

\end{small}


\begin{thebibliography}{99}

\bibitem{b1}
{\textit Gribov V. N., Lipatov L. N.}
Deep inelastic $e p$ scattering in perturbation theory //
Sov. J. Nucl. Phys. 1972. V. 15. P. 438-450.

\bibitem{b2}
{\textit Gribov V. N., Lipatov L. N.}
$e+ e-$ pair annihilation and deep inelastic $e p$ scattering in perturbation
theory //
Sov. J. Nucl. Phys. 1972. V. 15. P. 675-684.

\bibitem{b3}
{\textit Dokshitzer Yu. L.}
Calculation of the structure functions for deep inelastic
scattering and $e+ e-$ annihilation by perturbation theory in quantum
chromodynamics //
Sov. Phys. JETP 1977. V. 46. P.~641-653.

\bibitem{b4}
{\textit Altarelli G., Parisi G.}
Asymptotic freedom in parton language //
Nucl. Phys. B. 1977. V. 126. P.~298-318.

\bibitem{b5}
{\textit Forte S., Magnea L.}
Truncated moments of parton distributions //
Phys. Lett. B. 1999. V. 448. \mbox {P. 295-302.}

\bibitem{b6}
{\textit Forte S. et al.}
Evolution of truncated moments of singlet parton distributions //
Nucl. Phys. B. 2001. V. 594. P. 46-70.

\bibitem{b7}
{\textit Piccione A.}
Solving the Altarelli-Parisi equations with truncated moments //
Phys. Lett. B. 2001. V.~518. P.~207-213.

\bibitem{b8}
{\textit Forte S. et al.}
Determination of $\alpha_s$ from scaling violations of truncated moments
of structure functions~// 
Nucl. Phys. B. 2002. V. 643. P. 477-500.

\bibitem{b9}
{\textit Kotlorz D., Kotlorz A.}
Truncated moments of nonsinglet parton distributions
in the double logarithmic $ln^2 x$ approximation //
Acta Phys. Pol. B. 2004. V. 35. P. 705-721.

\bibitem{b10}
{\textit Kotlorz D., Kotlorz A.}
Evolution equations for truncated moments of the parton distributions //
Phys. Lett. B. 2007. V. 644. P. 284-287.

\bibitem{b11}
{\textit Kotlorz D., Kotlorz A.}
Evolution equations of the truncated moments of the
parton densities. A possible application //
Acta Phys. Pol. B. 2009. V. 40. P. 1661-1671.

\bibitem{b12}
{\textit Kotlorz D., Kotlorz A.}
Truncated Mellin moments: Useful relations and implications for the spin
structure function $g_2$ // 
Acta Phys. Pol. B 2011. V. 42, P. 1231-1246.

\bibitem{b13}
{\textit Psaker A. et al.}
Quark-hadron duality and truncated moments of nucleon structure functions //
Phys. Rev. C. 2008. V. 78. P. 025206.

\bibitem{b14}
{\textit Kumano S., Nagai T. -H.}
Comparison of numerical solutions for $Q^2$ evolution equations //
J. Comput. Phys. 2004. V. 201, P. 651-664 and ref. therein.

\bibitem{b15}
{\textit El-gendi S. E.}
Chebyshev solution of differential, integral and
integro-differential equations //
Comput. J. 1969. V. 12. P. 282-287.

\bibitem{b16}
{\textit Kwieci\'nski J., Maul M.}
Integral equation for spin dependent unintegrated parton distributions
incorporating double $ln^2(1/x)$ effects at low x // 
Phys. Rev. D 2003. V. 67, P. 034014.

\bibitem{b17}
{\textit HERMES Collaboration, Airapetian A. et al.}
Precise determination of the spin structure function $g_1$ of the proton,
deuteron and neutron //
Phys. Rev. D. 2007. V. 75. P. 012007.

\bibitem{b18}
{\textit COMPASS Collaboration, M. Alekseev M. et al.} 
The Polarised Valence Quark Distribution from semi-inclusive DIS  //
Phys. Lett. B. 2008. V. 660. P. 458-465.

\bibitem{b19}
{\textit Bl\"umlein B., B\"ottcher H.}
QCD Analysis of Polarized Deep Inelastic Data and Parton Distributions //
Nucl. Phys. B. 2002. V. 636. P. 225-263.

\bibitem{b20}
{\textit de Florian D., G.A. Navarro G. A., R. Sassot R.}
Sea quark and gluon polarization in the nucleon at NLO accuracy  //
Phys. Rev. D. 2005. V. 71. P. 094018.

\bibitem{BSR}
{\textit Bjorken J. D.}
Asymptotic sum rules at infinite momentum //
Phys. Rev. 1969. V. 179. P. 1547-1553.

\bibitem{COMPASS}
{\textit COMPASS Collaboration, M. Alekseev et al.}
The Spin-dependent Structure Function of the Proton $g_1^p$ and a Test
of the Bjorken Sum Rule // 
Phys. Lett. B. 2010. V. 690. P. 466-472.

\bibitem{b23}
{\textit Wandzura S., Wilczek F.}
Sum Rules for Spin Dependent Electroproduction:
Test of Relativistic Constituent Quarks //
Phys. Lett. B. 1977. V. 72. P. 195-198.

\bibitem{b24}
{\textit Burkhardt H., Cottingham W. N.}
Sum rules for forward virtual Compton scattering //
Ann. Phys. 1970. V. 56. P. 453-463.

\bibitem{b25}
{\textit Geyer B., Mueller D., Robaschik D.}
Evolution Kernels of Twist-3 Light-Ray Operators in Polarized Deep Inelastic Scattering  //
Nucl. Phys. B. Proc. Suppl. 1996. V. 51. P. 106-110.

\bibitem{b26}
{\textit Efremov A. V., Teryaev O. V., Leader E}
An Exact sum rule for transversely polarized DIS //
Phys. Rev. D. 1997. V. 55. P. 4307-4314.

\end{thebibliography}
\end{document}